\documentclass[aps,prl,twocolumn,superscriptaddress,showpacs,floatfix,10pt]{revtex4-1}
\usepackage{amsmath}
\usepackage{graphicx}
\usepackage{color}
\usepackage{epstopdf}
\usepackage{natbib}
\usepackage{xfrac}
\usepackage[normalem]{ulem}

\begin{document}
\title{Spin polarisation of ultrashort spin current pulses injected in semiconductors}

\author{M.~Battiato}
\email[]{marco.battiato@ifp.tuwien.ac.at}
\affiliation{Institute of Solid State Physics, Vienna University of Technology,  Vienna, Austria}

\begin{abstract}

Ultrashort spin current pulses have a great potential of becoming the carriers of information in future ultrafast spintronics. They present the outstanding property of an extremely compressed time profile, which can allow for the building up of spintronics operating at the unprecedented THz frequencies. The ultrashort spin pulses, however still lack other desirable features. For instance the spatial profile resembles more that of a spill rather than that of a spatially compressed pulse. Moreover the ultrashort spin current pulses can travel only across small distances in metals. The injection of the ultrashort spin pulses from the metallic ferromagnet, where they have to be generated, into a semiconductor is proposed as the first step to overcome both issues by allowing the exited electrons to propagate in a medium with few scatterings. However designing efficient interfaces for the injection is challenging due to practical constraints like chemical and structural stability. This work therefore expands the study of injection to a broader range of interfaces, and analyses how different metallic layers and semiconductors influence the amplitude, the spin polarisation and duration of the ultrashort pulses. This provides guidelines for the selection of efficient interfaces and, equally importantly, experimentally testable trends.

\end{abstract}

\date{\today}

\maketitle

The past years have seen the birth of a new field within the area of ultrafast dynamics: the ultrafast spin transport. The possibility of transporting spin in the femtosecond timescale was first highlighted experimentally \cite{Malinowski08}, even though at that time the driving microscopic process remained unclear. The insight arrived when superdiffusive spin transport \cite{Battiato10} was theoretically proposed as one of the mechanisms of the ultrafast demagnetisation \cite{Beaurepaire96}. Apart from pinpointing one of the  drivers of the ultrafast demagnetisation, that work clarified the microscopic mechanism of the ultrafast spin transport. The authors showed how laser excited electrons in a ferromagnetic metal undergo a strongly spin asymmetric diffusion, which in turns leads to a spin diffusion away from the irradiated region \cite{Battiato10,Battiato12,Battiato14}. The idea was experimentally verified by showing that spin polarised  electrons excited in an Fe film travelled through a Au substrate and could be measured at the opposite surface \cite{Melnikov11}. The microscopic understanding of the fs electronic and spin transport led to the prediction and verification of several important effects, like the possibility of triggering the ultrafast demagnetisation without a direct laser excitation \cite{Eschenlohrnatmater2013,Vodungbo2016}, the ultrafast increase of magnetisation \cite{Rudolf12}, or understanding of the origin of the THz emission associated to the ultrafast demagnetisation \cite{kampfrathnatnano2013}. Ultrafast spin transport has been proposed as the cause of the fs modification of nanoscale magnetic domains and domain walls \cite{Vodungbo12,Pfau12} and suggested as active during the all-optical switching \cite{Graves13}. The field of the fs spin diffusion has since attracted a vast attention 
\cite{SchellekensAPL13,Turgut13,Moisan14,vonKorffSchmising14,Schellekens14,Choi14,SchellekensPRB14,Wieczorek15,Choi15,Nenno16,Battiato16}. The precise amplitude of the contribution of superdiffusive spin transport to the total demagnetisation still remains a topic of active research \cite{Koopmans05,Carpene08,Krauss09,Zhang09,Koopmans10,Carva11,Carvanat11,Chimata12,Carva13,Wienholdt13,Mueller13,Locht15,Tows15,Krieger15}, and has been recently estimated to reach up to 30\% in the most favourable configuration \cite{SchellekensPRB14}.

\begin{figure}[th]
 \includegraphics[width=0.485\textwidth]{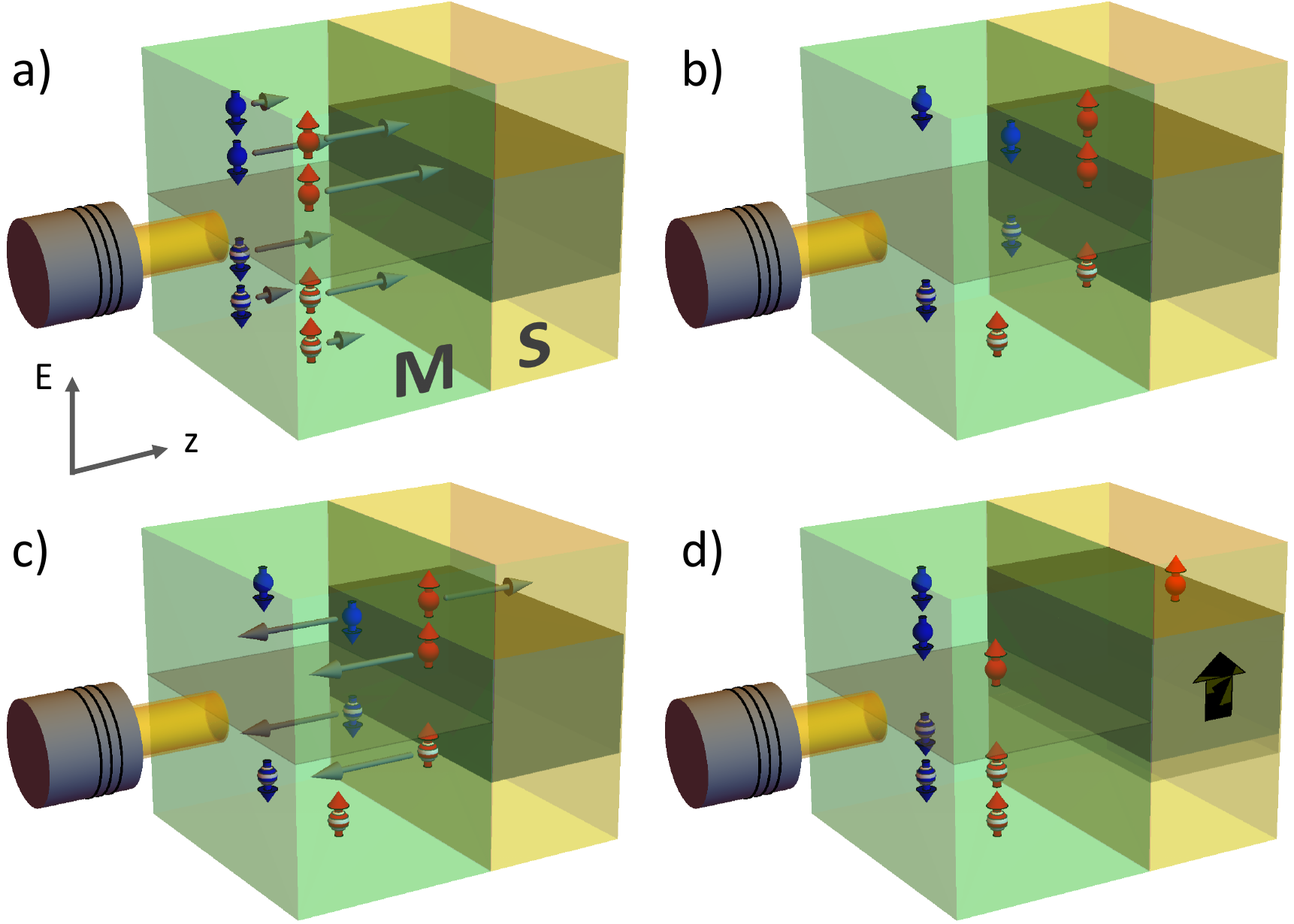}
 \caption{[Colour online] Schematic representation of the main processes happening at the metal/semiconductor (M/S) interface. The layer on the left (right) is the metal (semiconductor). The vertical axis represents the energy axis. The horizontal plane in the metallic layer is the Fermi level. Carriers above (below) it are electrons (holes). For simplicity only high and low energy carriers are shown. The shaded volume within the semiconductor represents the bandgap.  a) Laser-excited electrons and holes diffuse within the metallic layer with different efficiencies, represented by the length of the arrow pointing towards the semiconductor (notice that the calculated diffusion is in every direction). b) Carriers that diffuse efficiently reach more easily the M/S interface, while the other ones remain close to where they are excited. c) Electrons (holes) with energy above (below) the bandgap can cross into the semiconductor, while the others are reflected. d) The asymmetric injection leads to spin and charge accumulation within the semiconductor. The charge leads to the formation of electric fields that shift the position of the semiconductor's bandgap with respect to the metal's Fermi level.
}\label{fig:explan}
\end{figure}

Superdiffusive spin transport is a very promising candidate as a driver of information transport in the ultrafast timescale. The very short pulse length can in principle allow for the construction of THz spintronics, with no temporal overlap of the signals. Moreover spatial redistribution of magnetisation, does not incur in the same issues as the redistribution of charge. The latter creates electric fields that prevent the device to work at the desired extreme frequencies (see for instance Ref.~\onlinecite{Battiato16}).  
There is however a number of issues that still need to be solved in order to make these promising ultrashort spin current pulses usable. One of the critical problems is that the spatial profile resembles more a water spill (see for instance Fig.~6.b in Ref.~\onlinecite{Battiato12}) instead of a compact pulse. Another even more important problem is that the spin pulses propagate only up to few hundreds of nm in metals. The main reason is the very high number of scatterings in the metal. To circumvent this issue, in Ref.~\onlinecite{Battiato16} it has been proposed to produce the ultrashort spin pulse in ferromagnetic metals and then inject it in materials where the excited states experience dramatically lower number of scatterings, for instance semiconductors. This configuration gives the advantage of absorbing most of the laser power within a thin metallic layer (compared to irradiating the semiconductor directly) and creating high spin currents (due to the typically high magnetic moment of the ferromagnet). Moreover the short thermalisation time in the metal causes a quick switch off of the pulse leading to a very short pulse duration. The spin pulse is in this way able to propagate  within the semiconductor across technologically relevant distances (proportional to the mean free path in the used semiconductor). Moreover the injection in a semiconductor is the first step in producing a compact pulse (more on this below).

The problem of injection of spin in semiconductors is however not a new issue in materials science. The use of spin in modern electronics is already extremely successful \cite{Baibich88,Binasch89}. Integrating spintronics and semiconductor devices is of great interest due to the possibility of retaining the huge amount of existing expertise, as well as due to the  long spin lifetimes in semiconductors \cite{Kikkawa99}, or the possibility of avoiding Joule heating \cite{chumak15,Zutic11}. In spite of the fact that very promising  semiconductor-based spintronics for logic \cite{Awschalom07}, optical \cite{Endres13}, and thermoelectric \cite{Breton11,Jeon14} devices have been proposed, a full integration is still hindered by the challenge of efficiently injecting highly spin polarised currents in silicon \cite{Appelbaum11,Jansen12,Sverdlov15}. Direct injection through a ferromagnetic metal-silicon junction yields a negligible spin polarisation \cite{Schmidt00}. To circumvent the issues of the direct injection, different strategies are being investigated, as for instance injection of hot electrons \cite{Appelbaum07,Appelbaum11}, or the engineering of the ferromagnet-silicon contact by an additional insulating \cite{Jonker07,rashba00,Dash2009} or Schottky barrier \cite{Jansen12}. The proposal of injecting fs laser-excited carriers from a ferromagnetic metal into a semiconductor, bridges the fields of femtomagnetism and semiconductor spintronics.

However designing metal/semiconductor multilayers to be used for the ultrafast injection of spin faces a number of  practical problems. For instance the proposed Ni/Si structure \cite{Battiato16} has the critical issue of the formation of silicides at the interface, which, with their high bandgap, will prevent the injection. Unfortunately, the treatment of complex interfaces is still beyond the currently available theoretical possibilities. Therefore at this stage, to have a comparison with the theory, the materials need to be selected according to the strict requirement of producing an interface as close as possible to the idealised one described in the model. It is then critical to investigate the spin injection in a wide number of materials, to be able to accommodate growth requirements. This work analyses the impact of different choices for the metal and the semiconductor. Two ferromagnetic transitions metals (Ni and Fe) will be compared. As will be clarified below, within the investigated regime, the semiconductor's only relevant parameter is the bandgap. Finally, due to the competing needs of absorbing all the laser power within the metallic layer and preventing degradation of the spin pulse while traveling through it, the metallic layer thickness dependence will be studied as well. These results provide guidelines to design experiments and trends to be tested. 

Fig.~\ref{fig:explan} shows a schematic depiction of the process of injection. The laser pulse is absorbed mainly within the metallic layer (direct excitation of carriers in the semiconductor will be neglected in this work), and creates excited carriers  at different energies (Fig.~\ref{fig:explan}.a).  All the excited carriers will diffuse within the metal. As already highlighted  \cite{Battiato10,Battiato12,Battiato14} the diffusion cannot be described as standard diffusion and requires the use of a more complex model. The efficiency of the diffusion depends on the group velocities and scattering lifetimes, which are in turn dependent on the particle type (electron or hole), the spin channel and the energy. For a typical ferromagnetic transition metal, higher energy excited carriers diffuse less than low energy ones, majority electrons tend to have better transport properties than minority electrons, and both diffuse better than holes. An intuitive reason for this difference is that, in these materials, holes and minority electrons are mainly in localised d-states, while majority electrons in the more dispersive sp-states. The different effectiveness of the diffusion process for different carriers is pictorially depicted in Fig.~\ref{fig:explan}.a as arrows pointing towards the M/S interface, but it has to be understood that the carrier motion is that of an isotropic diffusion: the directionality towards the M/S interface comes from the fact that the metal/vacuum surface reflects back all the carriers, and that  the lateral diffusion towards the outside of the laser spot can be neglected due to the very high ratio between the size of the laser spot and the thickness of the film.

\begin{figure}[th]
 \includegraphics[width=0.485\textwidth]{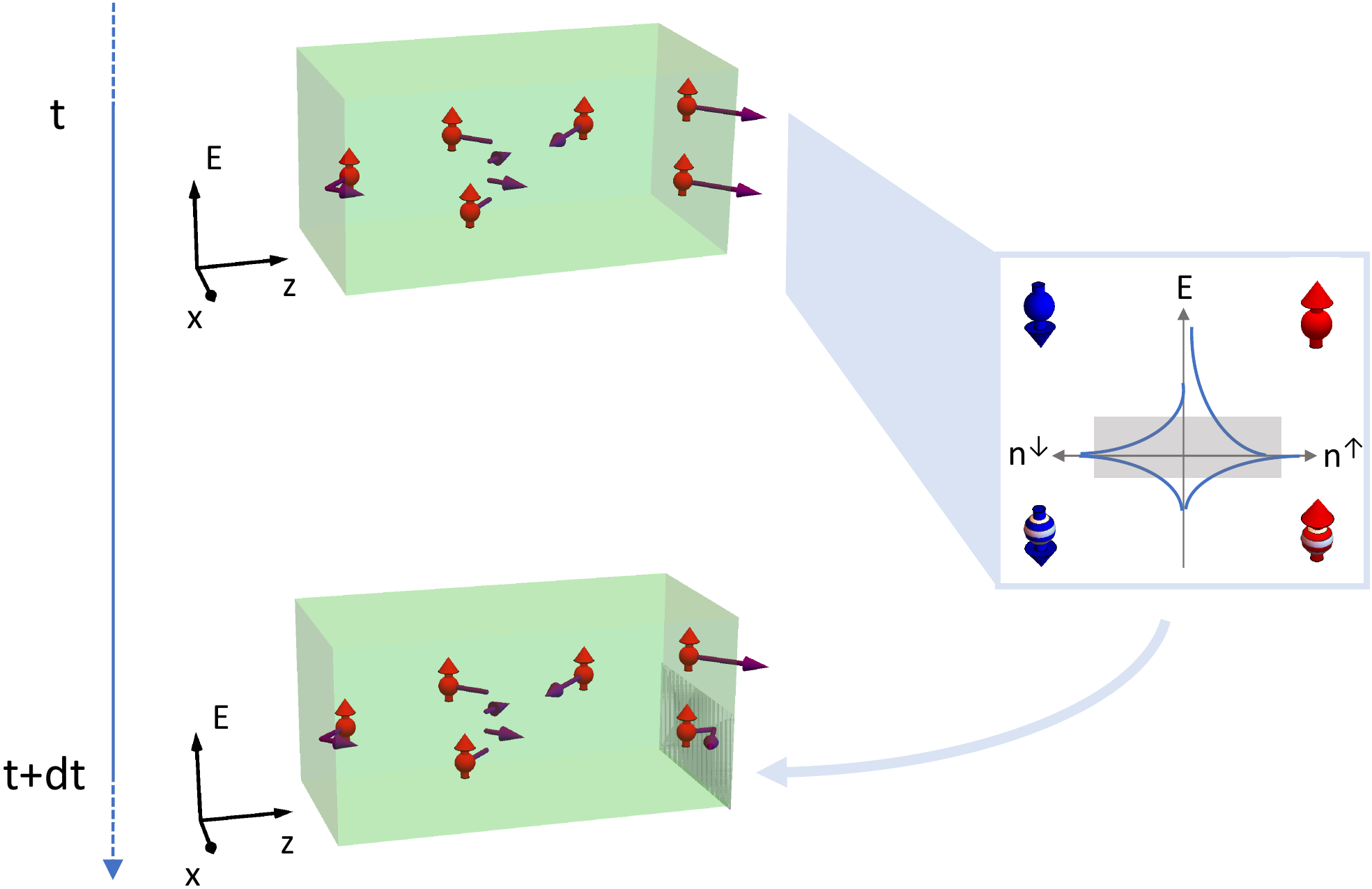}
 \caption{[Colour online] Structure of the time propagation in the solver in the high fluence regime. On the top left, the diffusion of the carriers is calculated. For simplicity only the diffusion of spin up electrons is showed, but all the species are calculated. The green box represents the metal. Only two spatial coordinates are shown, while the vertical exis represents the energy of the excited electron. From left to right electrons undergoing relevant modelled effects are shown: electron being reflected at the interface with the vacuum; two electrons at different energies scattering and changing their energies; one electron simply moving in space; and two electrons crossing the M/S interface at different energies. After one time step of the out-of-equilibrium transport in the metal is calculated, the energy, spin and type dependent flux at the M/S interface is extracted and the position of the semiconductor's bandgap is computed in order to ensure an uncharged total flux (centre right). Once the position of the bandgap is obtained, the transport outcome is corrected (bottom left) by allowing carriers outside of the bandgap to escape the metal, while carriers within it are reflected back. After this the process is repeated for the next time step.
}\label{fig:explanprogr}
\end{figure}

The asymmetry in the diffusion leads to an asymmetric flux of carriers reaching the metal/semiconductor (M/S) interface (Fig.~\ref{fig:explan}.b). Of the carriers reaching the interface, only those that have enough energy to overcome the bandgap can cross into the semiconductor, while the others will be reflected (Fig.~\ref{fig:explan}.c). The different injection in the two spin channels translates into a net spin current through the interface. Similarly the electron-hole asymmetry translates into a net charge crossing the M/S interface (Fig.~\ref{fig:explan}.d). For instance Fig.~\ref{fig:explan}.d shows a net negative charge injection in the semiconductor, which will charge the latter negatively and the metal positively (the sample as a whole remains neutral). The charged regions lead to the formation of electric fields, which will cause a (position dependent) shift of the semiconductor's bands with respect to those in the metal. This is shown in a simplified way in Fig.~\ref{fig:explan}.d, where the semiconductor's bandgap moves to higher energies. Before going in more details in the description of this effect, it is important to recognise three interesting qualitative behaviours. 1) In the linear regime, the intensity of the laser excitation does not affect the timescale of the thermalisation and diffusion within the metal, and simply rescales the density of excited carriers. This is however not true at the interface. The absolute amount of charge crossing the interface gives the amount of charge building up within an unit of time. The speed of charge accumulation is linked to the bandgap energy shift per unit of time. Therefore increasing the laser excitation leads to faster charging and faster motion of the bandgap position (compared to the timescale of the thermalisation/diffusion within the metal). 2) The motion of the bandgap at a given time will affect the transport at later times. In particular, in the case in Fig.~\ref{fig:explan}.d, the motion of the semiconductor's band structure acts towards a reduction (increase) of the amount of electrons (holes) injected at later times (in other words, the charging acts as a negative feedback mechanism trying to reduce the amount of charge injected). 3) In case the hypothetical case of steady state diffusion, the charging and consequently the shift of the bandgap will stop when the latter reaches a position that makes further total injection of excited carriers uncharged. However in reality the dynamics will be more complicated, since the energy distribution of the carriers that reach the M/S interface will change in time due to the thermalisation within the metal.

\begin{figure*}[th]
 \includegraphics[width=0.9\textwidth]{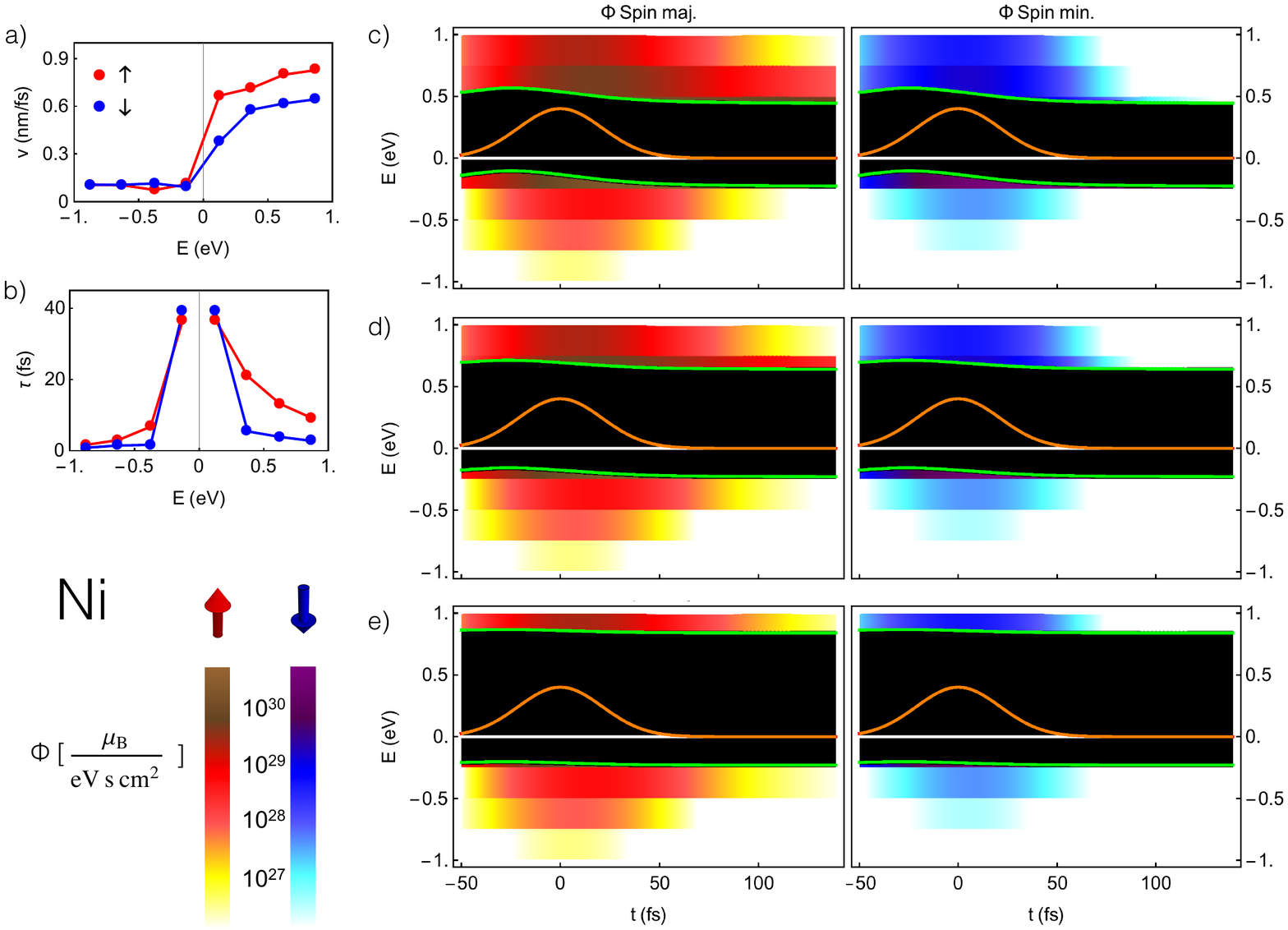}
 \caption{[Colour online] a) Energy-dependent group velocities for majority (up) and minority (down) excited electrons and holes in Ni. b) Excited electrons and holes lifetimes in Ni (including both e-e and e-ph scatterings \cite{Battiato16}). c-e) Energy- and time-dependent particle flux through the interface for a Ni(15 nm)/semiconductor sample resolved for spin majority (left) and minority (right) carriers for three different value of the semiconductor bandgap: c) 0.7 eV, d) 0.9 eV, and e) 1.1 eV. The intensity colour scale is in the bottom left corner. The green lines represent the dynamical position of the top of the valence and bottom of the conduction band. In the black energy range electrons and holes are reflected back into the metallic layer. The orange line represents the temporal profile of the pumping laser with a total fluence of 2.93 mJ/cm$^2$. 
}\label{fig:Ni}
\end{figure*}

The full modelling of the process  requires the description of the energy and spin dependent diffusion in the superdiffusive regime within both layers, with the explicit account of the scatterings. The electronic motion needs then to be coupled to the Poisson equation for the electric field. The complete dynamics is rather complicated (for instance in the general case the simplified picture in Fig.~\ref{fig:explan}.d, where the bandgap is simply rigidly shifted does not hold and the description of a position dependent shift is required \cite{Battiato16}). It was however showed that several approximations can be done to reduce the complexity. Within the metal the electric field generated by the charge diffusion is very efficiently screened by the motion of the carriers around the Fermi energy: this led to the derivation of the superdiffusive spin transport model \cite{Battiato10,Battiato12,Battiato14}. However that cannot be assumed within the semiconductor.  Ref.~\onlinecite{Battiato16}  showed that when a sufficiently high charge flux due to the excited carriers is injected in a semiconductor, the latter reacts almost instantly: the charge build-up at the interface and the consequent bandgap shift (Figs.~\ref{fig:explan}.c and \ref{fig:explan}.d) very quickly converge to the situation where only an uncharged flux is allowed to cross the M/S interface. This transient is much shorter than the thermalisation and transport dynamics within the metal and therefore negligible. This limit has been called high fluence regime. Please notice that effects that become important in the presence of strong electric fields, like the dynamical Franz-Keldysh effect, have not been included, and the bandgap size is taken as constant during the simulations. Moreover, in the high fluence regime, the charge within the semiconductor accumulates so close to the M/S interface that full account of the  position dependence of the bandgap shift is not required anymore and the effect of the electric field becomes equivalent to an effective position independent shift, as in Fig.~\ref{fig:explan}.d (see Ref.~\onlinecite{Battiato16} for more details).


The transport within the semiconductor can be further simplified by noticing that the scattering lifetimes in such materials are typically much longer than those in the metal (or at least one is interested in selecting  semiconductors with long lifetimes). We can therefore, as a first approximation, completely neglect scatterings in the semiconductor. In this case, once carriers coming from the metal are injected in a  semi-infinite semiconductor layer, they do not have the possibility of returning into the metal. This allows for a further simplification, as shown in Fig.~\ref{fig:explanprogr}. At each time step the diffusion of excited carriers in the metal can be computed by the superdiffusive spin transport model (top left image in Fig.~\ref{fig:explanprogr}). This gives rise to a flux at the M/S interface. The time-dependent position of the semiconductor's bandgap is then calculated at that time step in order to ensure that the integrated flux of electrons above the bandgap is equal to the integrated flux of holes below it, both summed over the two spin channels (centre right in Fig.~\ref{fig:explanprogr}). Once the instantaneous position of the bandgap is found, the motion of the particles at the M/S interface is corrected: the carriers being injected in the semiconductor are  removed from the calculation (since they will travel within the semiconductor, with no possibility of returning to the metallic layer), while the ones within the bandgap are reflected towards the metal (bottom right  in Fig.~\ref{fig:explanprogr}). The calculation proceeds by iterating the procedure above for each time step \cite{Battiato16}. Please notice that the further propagation of the injected carriers in the semiconductor is not addressed within the present work.

The process described above leads to a pure spin current injected through the M/S interface, with a very high spin polarisation \cite{Battiato16}. The reason is purely dynamical. The semiconductor bandgap allows only high energy electrons to cross the interface. However the out-of-equilibrium population in these states is strongly spin dependent. For instance in Ni majority electron-electron scattering lifetimes at high energies are longer than minority ones \cite{Zhukov06}. This leads to a quick depletion of the high energy minority states, even during the laser excitation, while the out-of-equilibrium population in the majority channel persists for a longer time. For this reason the high energy out-of-equilibrium electronic population in Ni is almost 100\% spin polarised (this is however not the case for the holes). Notice that as soon as the electronic system thermalises, this dynamical enhancement of the spin polarisation disappears: both spin channels become populated according to the same Fermi-Dirac distribution. The ultrafast injection of spin in semiconductors has therefore a huge advantage over, for instance, the steady state injection of hot electrons \cite{Appelbaum07,Appelbaum11} due to the purely dynamical effect of strongly spin-asymmetric relaxation times of the high energy electronic population.

\begin{figure*}[t]
 \includegraphics[width=0.9\textwidth]{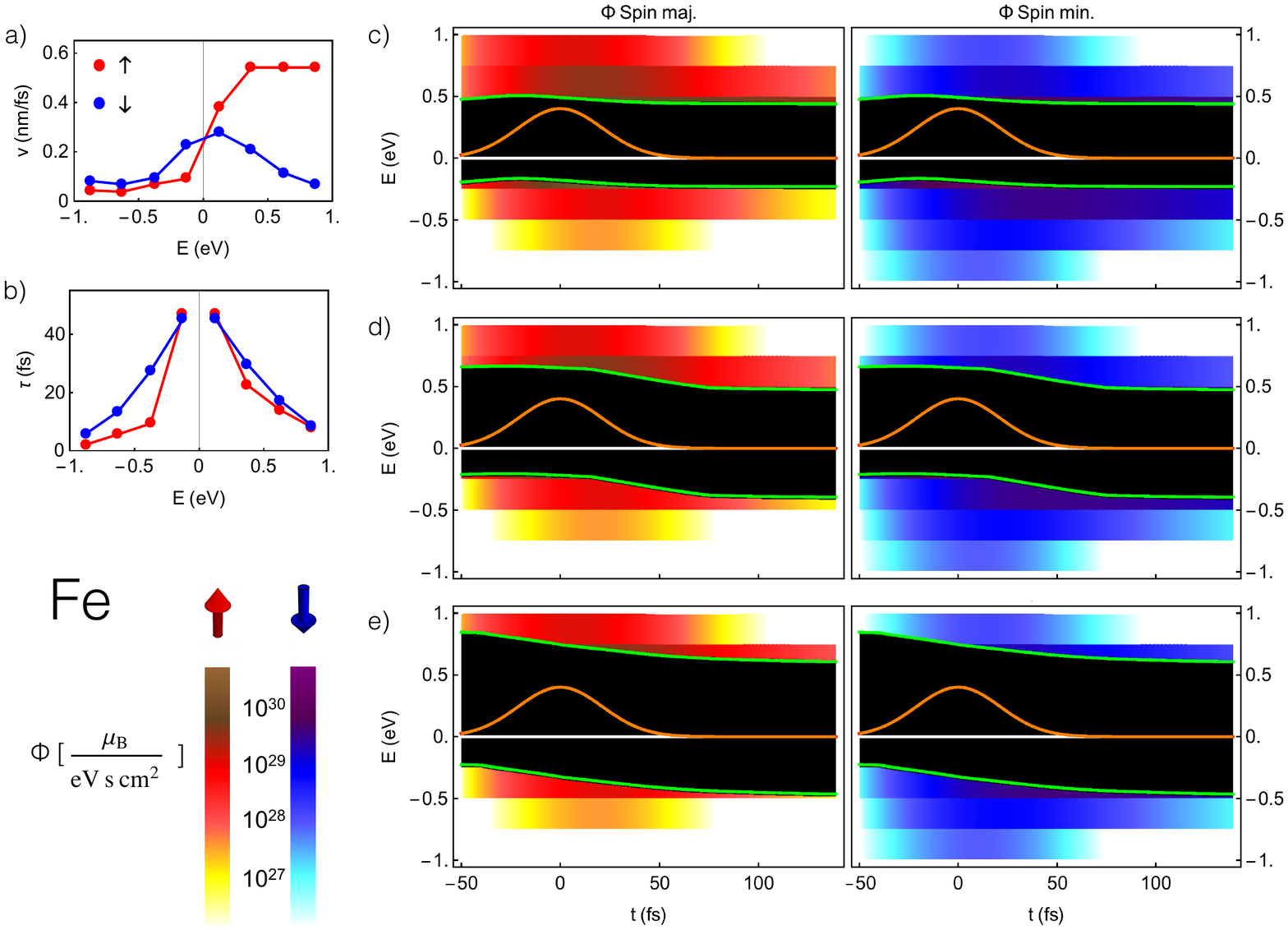}
 \caption{[Colour online] a) Energy-dependent group velocities for majority (up) and minority (down) excited electrons and holes in Fe. b) Excited electrons and holes lifetimes in Fe (including both e-e and e-ph scatterings  \cite{Battiato16}). c-e) Energy- and time-dependent particle flux through the interface for a Fe(15 nm)/semiconductor sample resolved for spin majority (left) and minority (right) carriers for three different value of the semiconductor bandgap: c) 0.7 eV, d) 0.9 eV, and e) 1.1 eV. The meaning of the lines is as in Fig.~\ref{fig:Ni}. The pumping laser total fluence is 2.93 mJ/cm$^2$
}\label{fig:Fe}
\end{figure*}

The energy- and spin-dependent velocities and lifetimes are displayed for Ni in Figs.~\ref{fig:Ni}.a and \ref{fig:Ni}.b, while for Fe in Figs.~\ref{fig:Fe}.a and \ref{fig:Fe}.b. These have been constructed from the density of states and \textit{ab initio} values \cite{Zhukov06} as described in Refs.~\onlinecite{Battiato16} and \onlinecite{BattiatoPhDThesis} (electron-phonon and electron-impurity scattering lifetimes are taken as 50fs for all energies and spin and for both materials). Excited majority electrons in Ni (which carry spin up) have slightly higher group velocities (Fig.~\ref{fig:Ni}.a) compared to minority electrons. However the strongest asymmetry comes from the scattering lifetimes (Fig.~\ref{fig:Ni}.b). We therefore expect the injection asymmetry to arise mainly from a strong imbalance of population at higher energies. The holes in both spin channels have instead very similar group velocities and lifetimes, leading to an almost unpolarised injection below the bandgap. The behaviour in Fe has important qualitative differences. The asymmetry for the electrons mainly comes from the velocities (Fig.~\ref{fig:Fe}.a) rather than the lifetimes (Fig.~\ref{fig:Fe}.b). Interestingly, in Fe the holes show a markedly different behaviour for the two spin channels. Minority holes (carrying spin up) have higher velocities and longer lifetimes compared to majority ones. Therefore we expect that an important part of the spin injection will be generated by hole diffusion.

We can now analyse in detail the case of a Ni layer with three different thicknesses (15 nm, 25 nm, and 35nm) deposited on a semiconducting substrate with three values of semiconducting bandgap 0.7 eV, 0.9 eV, and  1.1 eV. The system is excited by a 1 eV laser pulse (with a pulse width of 50 fs, and penetration depth of 10 nm). For simplicity the excitation profile, both for electrons and holes, is uniform in energy up to the maximum single photon excitation energy (multi photon excitations have been neglected). The energy scale, up to the maximum excitation, has been discretised in 8 energy channels (2 less than in Ref.~\onlinecite{Battiato16}), the spatial step has been taken as $1$nm, and the time step as $1$fs. The diffusion within the metallic layer is not shown, since, due to the relatively small thickness, the excited electron distribution does not show remarkable features: only a slow decrease of the intensity with the distance from the irradiated surface for the highest energy states (similarly to what shown in Ref.~\onlinecite{Battiato12}). In Figs.~\ref{fig:Ni}.c-e the spin-, energy-, and time-dependent flux at the metal/semiconductor interface is displayed for three values of the semiconductor bandgap (0.7 eV, 0.9 eV, and  1.1 eV respectively) for the Ni(15nm)/semiconductor configuration (similar pictures are obtained for the other two configurations with thicker Ni films and are not shown). Please notice the density is plotted as constant within each energy step, while in Ref.~\onlinecite{Battiato16} a linear interpolation was used (the figure appeared smoother than the real discretisation due to the fact that discrete changes in slope are not visible in a colour scale). 

Fig.~\ref{fig:Ni}.c provides a  view of the dynamical occupation of the highest energy states (please notice the logarithmic scale). Majority electrons reach the surface and the injection into the semiconductor persists for longer times, due to the long lifetimes. An out-of-equilibrium population at high energy states is present well after the laser pulse (whose temporal profile is displayed in orange in Figs.~\ref{fig:Ni}.c-e) has switched off. Minority electrons flux instead becomes negligible already few tens of fs after the laser excitation. The hole population quickly decays to low energies. The bandgap is therefore forced to high energies to allow enough holes to be injected in the semiconductor and thus compensate the electron flux.

\begin{figure}[t]
 \includegraphics[width=0.485\textwidth]{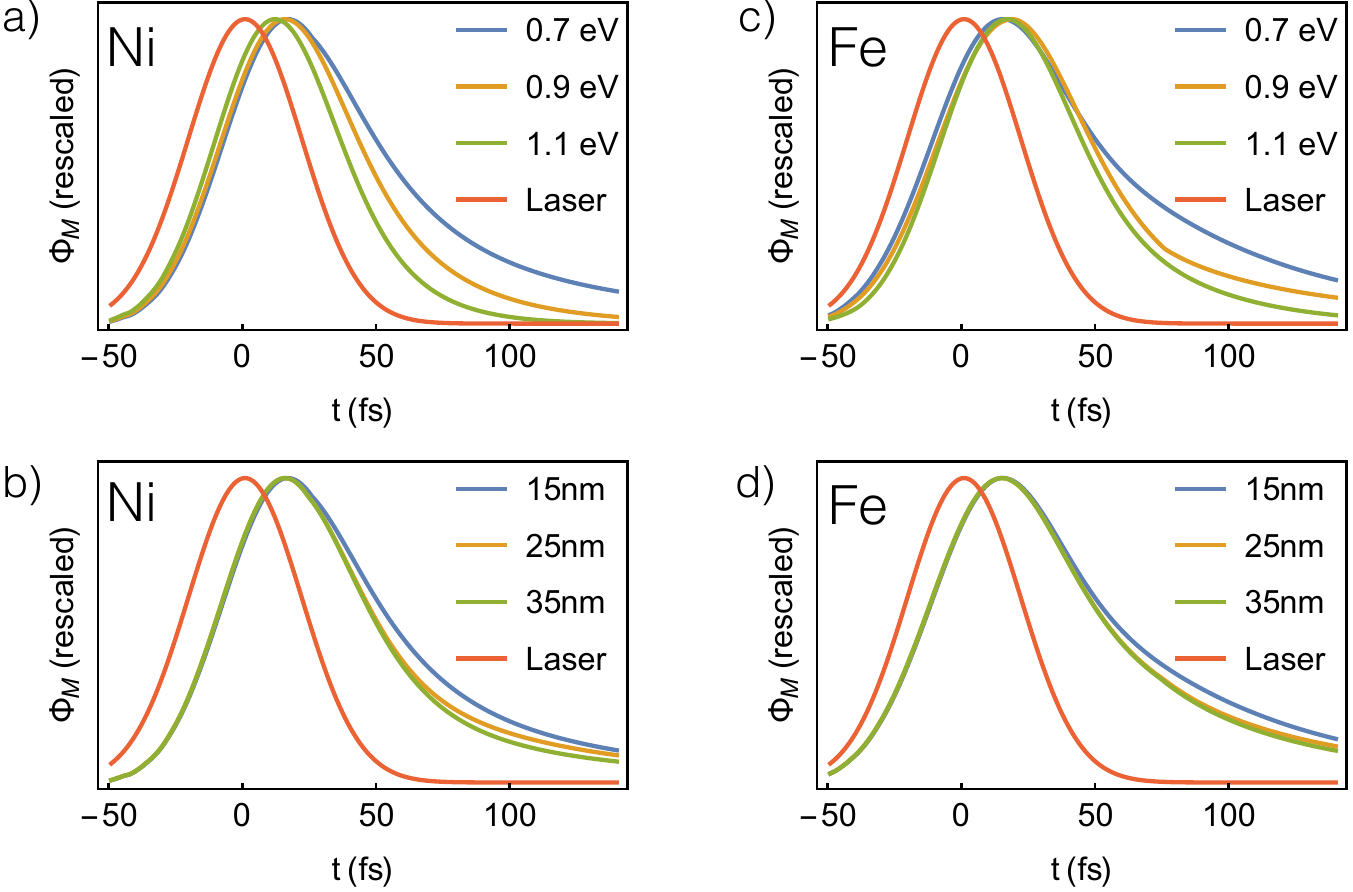}
 \caption{[Colour online] a) and c) Comparison of the time profile of the energy integrated magnetisation flux for the Ni(15 nm)/semiconductor and Fe(15 nm)/semiconductor samples for three different values of the bandgap. The temporal profile of the laser is provided for reference. All the curves have been rescaled to their maximum. b) and d) Comparison for the Ni/semiconductor and Fe/semiconductor samples with metallic layer of different thicknesses and semiconductor with the bandgap of 0.7 eV.}\label{fig:timedep}
\end{figure}


In Fig.~\ref{fig:timedep}.a the temporal shape of the magnetic flux through the interface has been compared for the three different semiconductor's bandgaps. In all the cases the maximum of the magnetisation flux is delayed with respect to the peak of the laser excitation. This is due to the competing interplay of three factors:  the fact that the number of excited electrons is proportional to the integrated intensity, the decay of population in the highest excited states and the multiplication of excited carriers upon electron-electron scattering. The most interesting feature of the comparison is that the magnetisation injection pulse duration increases with decreasing bandgap. For smaller bandgaps  lower energy states, with longer lifetimes, participate to the injection, allowing the injection to persist longer. Notice that this is an undesirable feature, since the aim is to achieve spin current pulses as compressed in time as possible. Fig.~\ref{fig:timedep}.b shows the weak dependence of the spin pulse temporal shape on the thickness of the metallic layer (showed only for a semiconductor substrate with a bandgap of 0.7 eV, similar dependences are obtained for larger bandgaps).

As mentioned before, the spin injection in semiconductors from Fe presents some interesting qualitative differences from that in Ni/semiconductor junctions. In Figs.~\ref{fig:Fe}.c-d the flux through the interface is plotted. We can observe that the asymmetry above the bandgap is not as strong as in the case of Ni, and excited electrons survive in the minority channel for a long time. Interestingly the hole channel is instead strongly asymmetric. Minority holes' long lifetimes, together with their relatively higher velocities, induce a hole flux which is strongly minority polarised. Given that minority holes transport spin up, this importantly contributes to the whole spin injection. The dependence of the temporal shape of the spin flux on the semiconductor's bandgap (Fig.~\ref{fig:timedep}.c) and metal's thickness (Fig.~\ref{fig:timedep}.d) show qualitative features very similar to the Ni case.

\begin{figure}[t]
 \includegraphics[width=0.485\textwidth]{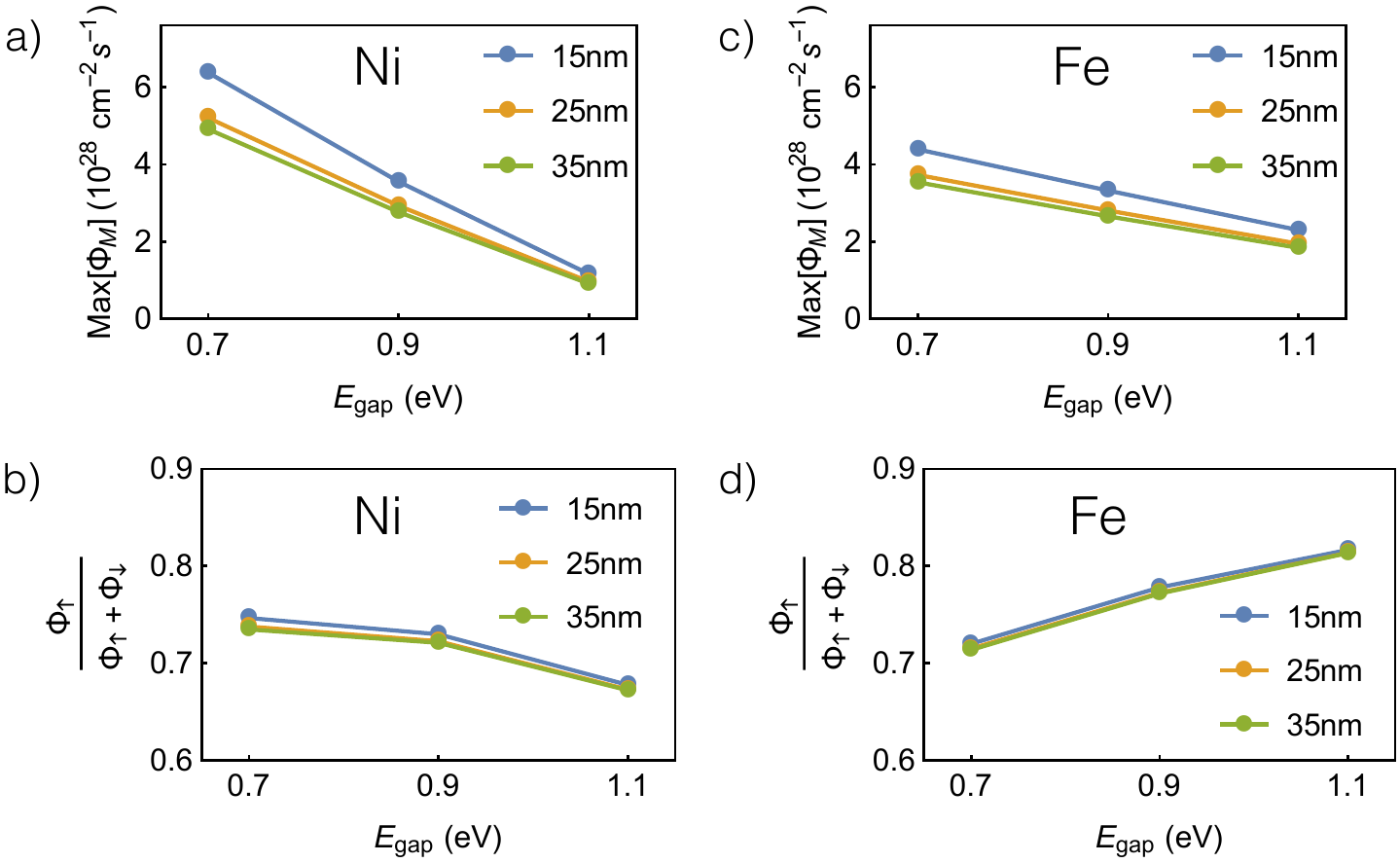}
 \caption{[Colour online] a) and c) Peak value of the magnetisation flux for Ni and Fe samples respectively in dependence of the metal layer's thickness and semiconductor's bandgap. b) and d) Fraction of flux originated by carriers carrying spin up for Ni and Fe samples respectively in dependence of the metal layer's thickness and semiconductor's bandgap.
 }\label{fig:character}
\end{figure}

The peak values of the magnetic flux are shown in Figs.~\ref{fig:character}.a and \ref{fig:character}.c for Ni and Fe respectively. The thickness of the metallic layer has a small influence (due to the relatively small thickness of the modelled metallic layers compared to the laser penetration depth). On the other hand the total intensity of the injected magnetisation across the metal/semiconductor interface shows the expected decrease with an increase of the size of the bandgap.  Interestingly in the case of Ni, the bandgap of the semiconductor strongly influences the amplitude of the injection, while the injection from the Fe does not decrease substantially with increasing bandgap. However the most remarkable difference between the two materials is that, while in the case of Ni lower bandgaps lead to higher spin polarisation (see Fig.~\ref{fig:character}.b), the opposite is true for Fe (see Fig.~\ref{fig:character}.d). 

The increase of spin polarisation with an increasing semiconductor's bandgap in the case of Fe can be understood by analysing the energy dependence of the velocities and lifetimes in Figs.~\ref{fig:Fe}.a and \ref{fig:Fe}.b. The lifetime asymmetry for the holes is the highest approximately below  $-0.25$eV, while closer to the Fermi energy the asymmetry falls sharply and even slightly reverses. In the case of small bandgap, the strong electron injection forces the bandgap to higher energies (Fig.~\ref{fig:Fe}.c). and allows for the injection of the spin unpolarised holes closer to the Fermi energy. Similarly, the higher the energy, the more majority polarised the electron transport is. Conversely in Ni the spin asymmetry of the excited electrons decreases with increasing energy (at least in the range of energies that are able to cross into the semiconductor). Given that the spin asymmetry of the holes plays a minor role in the spin injection, it is understandable that, with increasing bandgap and, therefore, selection of carriers only with higher energies, the spin polarisation drops.

After the spin pulse has been injected into the semiconductor, it can now travel across technologically relevant distances (depending on the mean free path in the semiconductor), in strong contrast to the transport within the metal. However it is worth highlighting that the injection of the spin into the semiconductor is only the first step in producing a compressed pulse, since the carriers crossing into the semiconductor will still not be injected only with velocities oriented perpendicularly to the interface. Ways to ensure it are to either use a semiconductor with anisotropic band structure or patterning to prevent lateral diffusion. Further studies are needed in this direction.

The description of the ultrafast injection of spin pulses in semiconductors proposed above has some limitations. As already mentioned, it can be applied only in the case of high carrier fluence from the metal. In the low fluence regime, 1) the charging of the M/S interface is no longer fast enough to neglect the transient within the semiconductor, 2) the extension of the charged region is not small enough to be treated as a simple reflecting surface, and 3) the injected carrier flux is not much larger than the charge that accumulates in the charged region. Apart from reconstructions of the interface, which are not accounted for in the simplified geometry used here, more subtle effects like energy-dependent partial reflection of the carriers above and below the bandgap are not included. This effect is expected to have increasing importance for thicker metallic layers (in a thin metallic layer carriers that are reflected at the M/S interface, can still traverse the metal, be reflected back at the metal/vacuum interface and attempt the injection again). A further simplification is that a spherically averaged band structure for the metals has been assumed. 

Concluding, this work  addresses how the ultrafast spin injection in semiconductors is influenced by the three most important variables in the high fluence regime: type of metallic ferromagnet, thickness of the metallic layer, and semiconductor's bandgap. This will help to select materials that have both a predicted strong injection, and  chemical and structural stability. It is also observed that the size of the bandgap of the semiconductor can be used both to control the temporal shape of the pulse and its spin polarisation. Interestingly Ni and Fe show differences in the characteristics of the injected spin current pulses. In particular Fe seems to be the most promising material, since the use of a wide bandgap semiconductor is predicted to lead to higher spin polarisation and more temporally compact spin current pulses, at the price of a relatively small loss in absolute amplitude of the spin flux. Finally this study provides testable trends to verify the  assumptions and approximations used in the theory.

\begin{acknowledgments}
The author is grateful to  K.~Held for very fruitful discussions. This work has been supported by the Austrian Science Fund (FWF) through Lise Meitner grant M1925-N28. 
\end{acknowledgments}


\end{document}